\documentclass[12pt]{article}

\begin{document}

\title{The Wahlquist metric cannot describe \\ an isolated rotating body}
\author{Michael Bradley$^1$, Gyula Fodor$^2$, Mattias Marklund$^3$ 
and Zolt{\'a}n Perj{\'e}s$^2$ \\
$1$ Department of Plasma Physics, Ume{\aa} University, S--901 87\\
Ume{\aa}, Sweden \\
$2$ KFKI Research Institute for Particle and Nuclear Physics,\\
H--1525, Budapest 114, P.O.B.\ 49, Hungary \\
$3$ Department of Mathematics and Applied Mathematics,\\
University of Cape Town, 7701 Rondebosch, Cape Town, South Africa}
\date{\today }
\maketitle

\begin{abstract}
It is proven that the Wahlquist perfect fluid space-time cannot be smoothly
joined to an exterior asymptotically flat vacuum region. The proof uses a 
power series expansion in the angular velocity, to a precision of the second 
order. In this approximation, the Wahlquist metric is a special case of the 
rotating Whittaker space-time. The exterior vacuum domain is treated in a 
like manner. We compute the conditions of matching at the possible boundary 
surface in both the interior and the vacuum domain. The conditions for 
matching the induced metrics and the extrinsic curvatures are mutually 
contradictory.
\end{abstract}

\section{Introduction}

   There is an embarrassing hiatus in the theory of rotating fluid bodies in 
general relativity. Whereas a large number of spherically symmetric static 
equilibrium states are known with their matching to the ambient 
vacuum \cite{Lake}, there is currently no global space-time model available 
for a rotating fluid body in an asymptotically flat vacuum. None of the few 
existing rotating fluid metrics could thus far been matched to a vacuum 
domain. An ominous case is that of the Wahlquist metric \cite{Wahl} which has 
been around and much studied for three decades. There is a parameter range 
within which the zero-pressure surface of the Wahlquist metric is prolate 
along the axis of rotation \cite{Hoenselaers}, and this indicates the action 
of an external force. Yet, no exact proof is known of the nonexistence of an 
asymptotically flat vacuum exterior. 

A convenient way of learning about the properties of fluids is to look at the 
slow-rotation limit. After all, if a property (for example, that of being 
possible to join a vacuum exterior) holds in an exact sense, then it is 
expected that the property holds in the slow-motion limit as well. In the
present work we develop an approximative scheme to the rotating Whittaker 
space-time. 
There are at least two reasons why this space-time is of interest for the 
research of the problem of relativistic fluid states: first, it is the static 
limit of the Wahlquist space-time; secondly, the equation of the rotation 
potential $\omega$ can be solved in terms of elementary functions.  

In this paper, the problem of matching of the rotating Whittaker metric to 
the vacuum exterior is investigated to a precision of quadratic order in the 
rotation parameter. 
It will be assumed that the fluid undergoes a circular rotation, that is, 
the velocity lies in the plane of the stationary ($\partial /\partial t$) 
and axial ($\partial/\partial \varphi $) Killing vectors. For rigid rotation, 
$u^{\varphi }=\Omega u^{t}$ where $%
\Omega $ is constant. We are using the approximation scheme for slow rotation, developed by Hartle 
\cite{Hart}, based on a power series expansion in the angular velocity   
$\Omega $ of the fluid. We keep at most second order terms. The metric has 
the form 
\newline
\begin{eqnarray}\label{ds2}
ds^{2} &=&[1+2h(r,\vartheta )]A(r)^{2}dt^{2}-[1+2m(r,\vartheta
)]B(r)^{2}dr^{2} \nonumber\\
&&-[1+2k(r,\vartheta )]C(r)^{2}\left\{ d\vartheta ^{2}+\sin ^{2}\vartheta
\left[ d\varphi +\left( \Omega -\omega (r)\right) dt\right] ^{2}\right\}
\end{eqnarray}
where $A, B$ and $C$ are the metric functions of the static configuration.  
The rotation potential $\omega$ is of order $ \Omega $ , and the functions  
$h$, $m$ and $k$ are of order $\Omega ^{2}$.

The $(t,\varphi )$ component of Einstein equation gives a second order
partial differential equation for $\omega $. The solution can be sought in
the form 
\begin{equation}
\omega (r,\vartheta )=\sum_{l=1}^{\infty }\omega _{l}(r)\left[ -\frac{1}{
\sin \vartheta }\frac{dP_{l}(\cos \vartheta )}{d\vartheta }\right] 
\end{equation}
The equations for the coefficients $\omega _{l}$ with different values of 
$l$ decouple. The functions $\omega _{l}(r)$ for $l>1$ cannot be
regular at infinity. Hence, by the
matching conditions it follows that the rotation potential is a function 
of the radial coordinate alone,  $\omega (r,\vartheta )=\omega (r)$
even in the fluid region. The potential $\omega$
satisfies a second-order ordinary linear differential equation.

We expand the second-order metric functions in Legendre polynomials, 
\begin{equation}
h(r,\vartheta )=\sum_{l=0}^{\infty }h_{l}(r)P_{l}(\cos \vartheta ) 
\end{equation}
and similarly for $m(r,\vartheta )$ and $k(r,\vartheta )$.
The quantities with different $l$  decouple once again.
The equations for $l>2$ do not include inhomogeneous terms with $\omega $, 
and it is asserted \cite{Hart} that the solution for $l>2$ does vanish as 
is the case in the static limit. Hence the Legendre expansion takes the form 
\begin{eqnarray}
h(r,\vartheta ) &=&h_{0}(r)+h_{2}(r)P_{2}(\cos \vartheta ) \\
m(r,\vartheta ) &=&m_{0}(r)+m_{2}(r)P_{2}(\cos \vartheta ) \\
k(r,\vartheta ) &=&k_{0}(r)+k_{2}(r)P_{2}(\cos \vartheta )
\end{eqnarray}
Hartle uses the freedom in the choice of radial coordinate to set $k_{0}=0.$
The second-order perturbations satisfy a set of inhomogeneous linear 
differential equations, with a driving term quadratic in the rotation 
function $\omega$.

 In Sec. 2, we review the form of the vacuum metric to the required accuracy.
In Sec. 3, we transform the slowly rotating Wahlquist metric to the Hartle 
form. What we obtain is a particular case of the rotating Whittaker fluid in 
this approximation. The matching surface ${\cal S}$ at $p=0$ is an ellipsoidal cylinder 
of rotation \cite{Hart}. We compute the normal $n$ and extrinsic curvature 
$K_{ab}$ of the embeddings of the surface ${\cal S}$ in both domains, in 
Sec. 4. These results are employed in Sec. 5, for setting up the junction 
conditions at ${\cal S}$. Here we conclude that the matching equations have 
no solution for the free constants of integration: they form an inconsistent 
set. In the light of earlier investigations on the nature of the matching 
equations of perfect fluids \cite{MS}, where it was found that the equations form an 
overdetermined set, our result is perhaps not that surprising. However, we 
argue as follows.  If the Wahlquist metric can be matched in general, then 
the matching would exist in the slow-rotation limit. Hence we conclude that 
the Wahlquist solution cannot be matched to an asymptotically flat region.

\section{The Vacuum Exterior}
 
 The metric of the ambient empty region 
to quadratic order has been computed in \cite{Hart}, albeit in a slightly 
different notation. The unperturbed metric is described by the 
Schwarzschild solution, 
\begin{equation}
A^{2}=\frac{1}{B^{2}}=1-\frac{2M}{r}\qquad ,\quad C=r\ .
\end{equation}
The perturbed metric is of the form (\ref{ds2}), with $ \Omega=0$
and
\begin{equation}
\omega =\frac{2aM}{r^{3}}\ , \qquad
\end{equation}
\begin{eqnarray}
h_{0} &=&\frac{1}{r-2M}\left( \frac{a^{2}M^{2}}{r^{3}}+\frac{r}{2M}
c_{2}\right) \\
m_{0} &=&\frac{1}{2M-r}\left( \frac{a^{2}M^{2}}{r^{3}}+c_{2}\right)  \\
h_{2} &=&3c_{1}r\left( 2M-r\right) \log \left( 1-\frac{2M}{r}\right) +a^{2}
\frac{M}{r^{4}}\left( M+r\right)  \nonumber\\
&&+2c_{1}\frac{M}{r}\left( 3r^{2}-6Mr-2M^{2}\right) \frac{r-M}{2M-r} \\
k_{2} &=&3c_{1}(r^{2}-2M^{2})\log \left( 1-\frac{2M}{r}\right) -a^{2}\frac{M
}{r^{4}}(2M+r) \nonumber\\
&&-2c_{1}\frac{M}{r}(2M^{2}-3Mr-3r^{2}) \\
m_{2} &=&6a^{2}\frac{M^{2}}{r^{4}}-h_{2}
\end{eqnarray}
To second order in the rotational parameter, the general slowly rotating 
solution is characterized by the mass parameter $M$, the
first order small rotation parameter $a$,  and the second order small 
constants $c_1$ and $c_2$.

A particular solution (the slowly rotating Kerr space-time) is given 
by $c_1=c_2=0$ and metric functions
\begin{eqnarray}
h_{0} &=&\frac{-a^{2}M^{2}}{r^{3}(2M-r)} \\[3mm]
h_{2} &=&\frac{a^{2}M(M+r)}{r^{4}} \\[3mm]
m_{0} &=&\frac{a^{2}M^{2}}{r^{3}(2M-r)} \\[3mm]
m_{2} &=&\frac{a^{2}M(5M-r)}{r^{4}} \\[3mm]
k_{2} &=&-\frac{a^{2}M(2M+r)}{r^{4}}\ .
\end{eqnarray}

\section{The Wahlquist solution}

The Wahlquist solution is given by the metric\cite{Wahl} 
\begin{eqnarray}
&&ds^{2}=f\left( dt-\tilde Ad\varphi \right) ^{2} \nonumber\\
&&\ \ -r_{0}^{\ 2}\left( \zeta ^{2}+\xi
^{2}\right) \left[ \frac{d\zeta ^{2}}{
\left( 1-\tilde k^{2}\zeta ^{2}\right) \tilde h_{1}}+
\frac{d\xi ^{2}}{\left( 1+\tilde k^{2}\xi ^{2}\right) \tilde h_{2}}
+\frac{\tilde h_{1} \tilde h_{2}}{\tilde h_{1}- \tilde h_{2}}d\varphi ^{2}\right] 
\end{eqnarray}
where 
\begin{eqnarray}
f &=&\frac{\tilde h_{1}-\tilde h_{2}}{\zeta ^{2}+\xi ^{2}}\quad ,\qquad 
\tilde A=r_{0}\left( 
\frac{\xi ^{2}\tilde h_{1}+\zeta ^{2}\tilde h_{2}}{\tilde h_{1}-\tilde h_{2}}-\xi _{A}^{\ 2}\right) \\
\tilde h_{1}\left( \zeta \right) &=&1+\zeta ^{2}+\frac{\zeta }{\kappa ^{2}}\left[
\zeta -\frac{1}{\tilde k}\sqrt{1-\tilde k^{2}\zeta ^{2}}
\arcsin \left( \tilde k\zeta \right)\right] \\
\tilde h_{2}\left( \xi \right) &=&1-\xi ^{2}-\frac{\xi }{\kappa ^{2}}\left[ \xi -
\frac{1}{\tilde k}\sqrt{1+\tilde k^{2}\xi ^{2}}\textrm{arcsinh}
\left( \tilde k\xi \right) \right]
\end{eqnarray}
The constant $\xi _{A}$ is defined by $\tilde h_{2}\left( \xi _{A}\right) =0$ . The
fluid velocity is proportional to $\partial /\partial t$ and the pressure
and density are 
\begin{equation}
p=\frac{1}{2}\mu _{0}\left( 1-\kappa ^{2}f\right) \quad ,\qquad \mu =\frac{1
}{2}\mu _{0}\left( 3\kappa ^{2}f-1\right) 
\end{equation}
Since the constants are related by the equation of state
\begin{equation}
\mu _{0}=\mu +3p=\frac{2\tilde k^{2}}{\kappa ^{2}r_{0}^{\ 2}} 
\end{equation}
we substitute 
\begin{equation}
\tilde k=\kappa r_{0}\sqrt{\frac{\mu_{0}}{2}} \ .
\end{equation} 

Making the coordinate transformation 
\begin{equation}
\zeta =r/r_{0}\quad ,\qquad \xi =\cos \Theta \left( 1+\frac{1}{12}
r_{0}^{2}\mu _{0}\right) 
\end{equation}
in the limit $r_{0}\rightarrow 0$ we get Whittaker's spherically symmetric
static solution\cite{Whit}. Making a further coordinate transformation 
\begin{equation}
\sin X=\kappa r \sqrt{\frac{\mu_{0}}{2}} \ ,
\end{equation} 
we get a somewhat simpler form for the metric in the 
$r_{0}\rightarrow 0$ limit 
\begin{equation}
ds^{2}=f_{0}dt^{2}-\frac{2}{\mu _{0}\kappa ^{2}}\left[ \frac{dX^{2}}{f_{0}}
+\sin ^{2}X\left( d\Theta ^{2}+\sin ^{2}\Theta d\varphi ^{2}\right) \right] 
\end{equation} 
where 
\begin{equation}
f_{0}=1+\frac{1}{\kappa ^{2}}\left( 1-X\cot X\right) \ . 
\end{equation}
Since the vanishing pressure surface is at $\kappa ^{2}f_{0}=1$ , from the
positivity of the pressure follows that the
constant $\kappa $ satisfies $0<\kappa <1,$ and the $X$ coordinate has the
range $0<X<X_{s}<\frac{\pi }{2}$ .

In the slowly rotating limit, if we keep only linear terms in $r_{0}$ , the
only change will be that we write $\left( d\varphi -\omega dt\right)
^{2}$ in place of $d\varphi ^{2}$ in the metric. To this accuracy we have $\xi _{A}=1$.
Thus the non-diagonal term in the metric is proportional to $\sin ^{2}\Theta $
, we get 
\begin{equation}
\omega =\frac{\mu _{0}r_{0}}{2\sin ^{2}X}\left( 1-X\cot X\right) 
\end{equation}

Keeping second-order terms in $r_{0}$, the metric has the form 
\begin{eqnarray}
ds^{2} &=&f_{0}(1+2h)dt^{2}-\frac{2}{\mu _{0}\kappa ^{2}}\frac{1+2m}{f_{0}}%
dX^{2} \nonumber\\ \label{dsW}
&&-\frac{2}{\mu _{0}\kappa ^{2}}\sin ^{2}X\left[ (1+2k)d\Theta ^{2}+\sin
^{2}\Theta (1+2n)\left( d\varphi -\omega dt\right) ^{2}\right]
\end{eqnarray}
where 
\begin{eqnarray}
h &=&{\frac{\cos^{2}\Theta}{2\sin^{2}X}(X\cot X-1)\mu_0r_{0}^{2}}
\\
m &=&{\frac{\kappa ^{2}\sin X
-\cos ^{2}\Theta (\kappa ^{2}\sin X+ \sin X-X\cos X)
}{\sin X(\kappa ^{2}\sin X+\sin X-X\cos X)^{2}}
\kappa ^{2}r_{0}^{2}} \\
k &=&-{\frac{1}{3\kappa ^{2}}}\left[3\kappa^2\cos^2 X\cos^2\Theta
+\sin^2 X\left(1+\cos^2\Theta\right)\right]r_0^2 \\
n &=& \frac{\sin ^{2}\Theta }{3\kappa ^{2}\sin X}
\left[3\sin^2\Theta\left(\sin X-X\cos X\right) 
\right.\nonumber\\ &&\hspace*{22mm} \left.
+\cos ^{2}\Theta \sin^{3}X-3\kappa ^{2}\sin X\right]r_{0}^{2}
\end{eqnarray}

To get into Hartle`s coordinates, $n=k$ and $k_{0}=0$, we make an
infinitesimal coordinate transformation 
\begin{equation}
X=x+\alpha (x,\vartheta )r_{0}^{2}\quad ,\quad \Theta =\vartheta +\beta
(x,\vartheta )\sin \vartheta r_{0}^{2} 
\end{equation}
where 
\begin{eqnarray}
\alpha (x,\vartheta ) &=&\alpha _{0}+\alpha _{2}\sin ^{2}\vartheta \\
\beta (x,\vartheta ) &=&\beta _{1}\cos \vartheta
\end{eqnarray}
and 
\begin{eqnarray}
\alpha _{0} &=&\frac{-\mu _{0}}{12\kappa ^{2}\cos x\sin ^{3}x}%
\{[3x^{2}-(2x^{2}-3)\sin ^{2}x]\cos ^{2}x \\
&&-x\cos x\sin x(6-5\sin ^{2}x)+\kappa ^{2}\sin ^{4}x\} \\[3mm]
\alpha _{2} &=&\frac{-\mu _{0}}{8\kappa ^{2}\cos x\sin ^{3}x}[x\cos x\sin
x(\kappa ^{2}+6-5\sin ^{2}x) \\
&&+\left( 2\kappa ^{4}-\kappa ^{2}+2x^{2}-3\right) \sin ^{2}x\cos
^{2}x-3x^{2}\cos ^{2}x] \\[3mm]
\beta _{1} &=&\frac{-\mu _{0}}{12\sin ^{3}x}\left[ 3(x\cos x+\kappa ^{2}\sin
x-\sin x)-(3\kappa ^{2}-2)\sin ^{3}x\right]
\end{eqnarray}
To avoid the appearance of logarithm terms in $\vartheta $ we rescale the
angular coordinate by a small constant factor 
\begin{equation}
\varphi \longrightarrow \left[ 1+\frac{1}{4}r_{0}^{2}\mu _{0}\left( 1-\kappa
^{2}\right) \right] \varphi 
\end{equation}
This will also ensure that there will be no conical singularity at the axis.
We also rescale the time coordinate $t$ to cancel a second order small 
constant term in $h$.

The metric (\ref{dsW}) takes the form (\ref{ds2})
\begin{eqnarray} 
ds^{2} &=&f_{0}(1+2h)dt^{2}
-2\frac{1+2m}{\mu _{0}\kappa ^{2}f_{0}}dx^{2} \nonumber \\ &&
-\frac{2}{\mu _{0}\kappa ^{2}}\sin ^{2}x (1+2k)\left[d\vartheta ^{2}+\sin
^{2}\vartheta\left( d\varphi -\omega dt\right) ^{2}\right] \label{dsw}
\end{eqnarray}
with
\begin{equation}
f_{0}=1+\frac{1}{\kappa ^{2}}\left( 1-x\cot x\right) 
\end{equation}
and
\begin{equation}
\omega =\frac{\mu _{0}r_{0}}{2\sin ^{2}x}\left( 1-x\cot x\right)  \ .
\end{equation}
For the second order small quantities we obtain
\begin{eqnarray}\label{secord}
h_{0} &=& x\frac{5\sin^{2}x-6-2\cos ^{2}x\sin ^{2}x}
{24\left[ \left( \kappa ^{2}+1\right) \sin {x}-{x
}\cos {x}\right] \sin ^{2}{x}\cos {x}} \mu_0r_{0}^{2} \nonumber\\
&&-\frac{4x^{2}\sin ^{2}x-3\sin ^{2}x-3x^{2}-4\sin ^{4}x}{24\left[ \left(
\kappa ^{2}+1\right) \sin {x}-{x}\cos {x}\right] \sin ^{3}{x}}\mu_0
r_{0}^{2}  \\
h_{2} &=&\left( x\cos {x}\frac{\sin ^{2}x+3\kappa ^{2}-3}{6\sin ^{3}x}+\frac{
\kappa ^{2}}{6}
\right. \nonumber \\ && \left. \hspace*{5mm}
-\frac{1}{4}+\frac{3-6\kappa ^{2}-2x^{2}}{12\sin ^{2}x}+\frac{
x^{2}}{4\sin ^{4}x}\right) \frac{\mu_{0}r_{0}^{2}}{2\kappa ^{2}} \\
m_{0} &=&-\left[ \frac{\sin ^{2}x}{12\cos ^{2}x}+x\frac{x\cos ^{3}x-2\cos
^{2}x\sin {x}-3\sin ^{3}x}{8\left[ (\kappa ^{2}+1)\sin {x}-{x}\cos {x}
\right] \sin ^{3}x}\cos x\right] \mu_{0}r_{0}^{2}  \nonumber \\
&&+\frac{x\sin ^{3}x+(x^{2}-7)\cos {x}\sin ^{2}x-(2x^{2}+3)\cos ^{3}x}{
24\left[ (\kappa ^{2}+1)\sin {x}-{x}\cos {x}\right] \cos x\sin x}\mu_0
r_{0}^{2} \\
m_{2} &=&\left( \frac{15x^{2}}{\sin ^{4}x}-6x\cos {x}\frac{\kappa
^{2}+5-\sin ^{2}x}{\sin ^{3}x}
\right. \nonumber \\ && \left. \hspace*{5mm}
-2\kappa ^{2}-7-\frac{14x^{2}-15-6\kappa ^{2}}{
\sin ^{2}x}\right) \frac{\mu_{0}r_{0}^{2}}{24\kappa ^{2}}  \nonumber \\
k_{2} &=&\left( \frac{5x^{2}-3+3\kappa ^{2}}{\sin ^{2}x}
-2x^{2}+3-\kappa ^{2}
\right. \nonumber \\ && \left. \hspace*{5mm}
-x\cos {x}\frac{3\kappa ^{2}-6+5\sin ^{2}x}{\sin ^{3}x}
-\frac{3x^{2}}{
\sin ^{4}x}\right) \frac{\mu_{0}r_{0}^{2}}{12\kappa ^{2}}  
\end{eqnarray}
All these second order quantities go to zero as $x \rightarrow 0$, which 
shows that the center is regular to the required order. 

The pressure is $p=p_{0}+r_{0}^{2}\left[ p_{20}+p_{22}P_{2}\left( \cos
\vartheta \right) \right] $ where 
\begin{eqnarray}
p_{0} &=&\frac{\mu_{0}}{2}\left( x\cot x-\kappa ^{2}\right) 
\label{eqp0} \\
p_{20} &=&\frac{\mu_{0}^{2}}{24\sin ^{4}x\cos x}\left[ x\sin x\left( 3\sin
^{2}x-2\right) 
\right. \nonumber \\ && \left. 
+\left( 2\kappa ^{2}\sin ^{4}x-2\sin ^{4}x+\sin
^{2}x+x^{2}\right) \cos x\right] \label{eqp20}\\
p_{22} &=&-\frac{2\kappa ^{4}\mu_{0}^{2}}{12\sin ^{2}x}
+\frac{\mu_{0}^{2}}{24\kappa ^{2}\sin ^{5}x}
\left[ x\cos x-\left( \kappa ^{2}+1\right) \sin x\right] 
\nonumber \\ &&  
\left[ \left( 2\kappa ^{2}-3\right) \sin ^{4}x
 +2x\sin x\cos x\left( \kappa ^{2}-3+\sin ^{2}x\right) 
\right. \nonumber \\ && \left. \hspace*{2mm} 
 -\left(
2x^{2}-3+4\kappa ^{4}+2\kappa ^{2}\right) \sin ^{2}x+3x^{2}\right]
\label{eqp22}
\end{eqnarray}

\section{The matching surface}

In the fluid region, the matching surface ${\cal S}$ is defined by the
condition of vanishing pressure, $p=0.$ In the limit of no rotation, the
matching surface is the history of the sphere $x=x_{1}$ statisfying
\begin{equation}
x_1\cot x_1=\kappa^2 \ .  \label{x1kappa}
\end{equation}
For slow rotation the equation of
the matching surface ${\cal S}$ is 
\begin{equation}\label{SW}
x=x_{1}+r_{0}^{2}\xi 
\end{equation}
with 
\begin{equation}
\xi =-\left[\frac{p_{20}+p_{22}P_2(\cos \vartheta) 
}{p_{0}^{\prime }}\right]
_{\mid \;x=x_{1}}
\end{equation}
where we denote $d/dx$ by a prime. Substituting (\ref{eqp0}) -- 
(\ref{eqp22}) and (\ref{x1kappa}), we get that 
$\xi=\xi_0+\xi_2 P_2(\cos\vartheta)$, where the constants
$\xi_0$ and $\xi_2$  are defined by
\begin{eqnarray}
\xi_0&=&
\mu_0 {\frac {  
\kappa^{10}-2 \kappa ^{8}+2 x_1^{2}\kappa^{6}+\kappa^{6}
+x_1^{2}\kappa^{4}+x_1^{4}\kappa^{2}-x_1^{2}\kappa^{2}+x_1^{4}
}{12x_1 \kappa^{2}\left(
\kappa^{4}-\kappa^{2}+x_1^{2}
\right )}}
 \\
\xi_2&=&
-\mu_0{\frac {
4 \kappa^{10}+\kappa^{8}+4 x_1^{2}\kappa^{6}-8 \kappa^{6}
+2 x_1^{2}\kappa^{4}+3 \kappa^{4}-4 x_1^{2}\kappa^{2}+x_1^{4}
}{12x_1 \kappa^{2}\left (
\kappa^{4}-\kappa^{2}+x_1^{2}\right )}}
\end{eqnarray}

In the vacuum exterior region, suitable hypersurfaces for matching 
are determined by the condition\cite{Roos}
\begin{equation}
\tilde \Omega ^{2}g_{\varphi \varphi }+2\tilde \Omega g_{\varphi t}
+g_{tt}=1-\tilde C
\label{roosc}
\end{equation}
where $\tilde \Omega $ and $\tilde C$ are constants.
In the limit of no rotation, the
matching surface is the history of the sphere $r=r_{1}.$ For slow rotation the deformation
of the surface is described by
\begin{equation}\label{SV}
r=r_{1}+a^{2}\chi 
\end{equation}
where $\chi $ is a function of $\vartheta $ and $a$ is the small rotational
parameter. Substituting into (\ref{roosc}) and keeping only second order
small terms in the rotational parameter, we get
\begin{equation}
\chi =\chi _{0}+\chi _{2}P_2(\cos \vartheta) 
\end{equation}
where $\chi _{0}$ and $\chi _{2}$ are constants to be determined by the
matching conditions.

The function $y=x-r_0^2 \xi$ characterizes the constant pressure
hypersurfaces, with $y=x_1$ on the matching surface. Denoting the 
coordinates as $x^a_{(W)}=(t,x,\vartheta,\varphi)$, the normal one-form
has the components
\begin{equation}
n_a^{(W)}=\left(0,1,3r_0^2\xi_2\sin\vartheta\cos\vartheta,0\right)
\sqrt{-g^{(W)}_{11}} 
\end{equation}
where $g^{(W)}_{ab}$ denotes the metric components in the Wahlquist 
region. Using the cooordinates $x^a_{(V)}=(t,r,\vartheta,\varphi)$ and 
introducing the notation $g^{(V)}_{ab}$ for the metric in the vacuum 
region, the normal form of  the possible matching surfaces is
\begin{equation}
n_a^{(V)}=\left(0,1,3a^2\chi_2\sin\vartheta\cos\vartheta,0\right)
\sqrt{-g^{(V)}_{11}} \ .
\end{equation}

The extrinsic curvature $K=K_{ab}dx^{a}dx^{b}\vert_{\cal S}$ of the surface ${\cal S}$ is defined in terms of 
the normal $n_a$ and the projector $h_{ab}=g_{ab}+n_an_b$ where $g_{ab}$ 
are the components of (\ref{ds2}) and the coordinates are restricted to  ${\cal S}$. The quantities $K_{ab}$ are given by
$K_{ab} =h_{a}\!^{c}h_{b}\!^{d}n_{(c;d)} $.
From the expression 
\begin{equation}
K_{ab} =n_{\left( a,b\right) }-n^{r}\left( g_{\left( a|r|,b\right)
}-\frac{1}{2}
g_{ab,r}\right) +n_{(a}n^{r}\left( n_{b),r}-n_{\left| r\right| ,b)}\right)
\end{equation}
for the extrinsic curvature we obtain 
\begin{eqnarray}
K_{00} &=&\frac{1}{2}g_{00,1}n^{1}\ , \\
K_{03} &=&\frac{1}{2}g_{03,1}n^{1}\ , \\
K_{12} &=&-\frac{1}{2}g_{22,1}n^{2}\ , \\
K_{22} &=&\frac{1}{2}g_{22,1}n^{1}+n_{2,2}\ , \\
K_{33} &=&\frac{1}{2}g_{33,1}n^{1}+n_{2}\sin \vartheta \cos \vartheta \ .
\end{eqnarray}
On the matching surface ${\cal S}$, given by Eqs. (\ref {SW}) and (\ref{SV}) respectively,
the parts $K_{11}dr^2$ and  $K_{12}dr d\theta$ of the extrinsic curvature will both be
of fourth order in the expansion parameters, hence dropped.

\section{Junction conditions}
 In this section, we search for isometric embeddings of the matching surface 
${\cal S}$ in the vacuum and Wahlquist domains, respectively. We equate with 
each other the respective induced extrinsic curvatures $K_{(V)}$ and 
$K_{(W)}$ of 
${\cal S}$, in the vacuum and in the Wahlquist region.
Hence, in terms of the induced metric $ds^2\vert_{\cal S}$ the equations of 
matching are
\begin{equation}\label{matchcd}
ds^2_{ (V)}\vert_{\cal S}=ds^2_{ (W)}\vert_{\cal S}\qquad
K_{(V)}\vert_{\cal S}=K_{(W)}\vert_{\cal S}\ .
\end{equation}
The matching to zero-order in the rotation parameter takes place on the 
cylinder which is the product, $S^2\times R$, of the metric two-sphere and the time. 
We take advantage of the freedom in taking constant 
linear combinations of the time and azimuthal coordinates. 
We  apply a rigid rotation in the fluid region by setting 
$\varphi\to\varphi+\Omega t$ where $\Omega$ is a constant. 
Then we re-scale the interior time coordinate $t\in R$ by 
$t\to c_4(1+r_0^2c_3)t$ with further constants $c_3$ and $c_4$ to be
determined from the maching conditions.

From the zero-order matching conditions we get the following relations:
\begin{eqnarray}
M&=&\frac{r_1}{2\kappa^2}(\kappa^2-\cos^2x_1)\\
r_1&=&\frac{2^{1/2}}{\kappa{\mu_0}^{1/2}}\sin x_1\\
c_4&=&\cos x_1
\end{eqnarray}

We next perform the matching to first order, with the matching surface still 
being the product $S^2\times R$. The matching equations are the $(t,\varphi)$ 
components of Eqs. (\ref{matchcd}) from which we get the parameter values
\begin{eqnarray}
\Omega&=&\frac{\mu_0x_1 r_0}{6\sin x_1\cos x_1}\\
a&=&\frac{r_0}{3\cos x_1}\,
    \frac{2x_1\cos^2x_1-3\sin x_1\cos x_1+x_1}{\sin x_1\cos x_1-x_1}\ .
\end{eqnarray}

 To second order in the rotation parameter, the matching surface $\cal{S}$ is 
an ellipsoidal cylinder characterized by the embedding conditions (\ref{SW}) and 
(\ref{SV}). The values of the metric coefficients and their derivatives on 
$\cal{S}$ are given by a power series expansion in $r_0$ in the fluid and 
in $a$ in the vacuum regions, respectively. In the vacuum, the nonvanishing 
components of the normal are
\begin{equation}n^1_{(V)}=-\frac{\cos x_1}{\kappa}(1-m_{(V)}) \qquad
n_{2(V)}=3\cos\theta\sin\theta\frac{\kappa a^2\chi_2}{\cos x_1}
\end{equation}
In the interior, we have
\begin{equation}n^1_{(W)}=-\left(\frac{\mu_0}{2}\right)^{1/2}(1-m_{(W)})\qquad
n_{2(W)}=3\cos\theta\sin\theta\left(\frac{2}{\mu_0}\right)^{1/2}r_0^2\xi_2\ .
\end{equation}

Substituting in the matching conditions (\ref{matchcd}), and Taylor 
expanding in the powers of the rotation parameter, we get a set of
lengthy linear equations. 
We need to solve these for the parameters  $\chi_0$, $\chi_2$, $c_1$, $c_2$
and $c_3$ in terms of $r_0$, $x_1$ and $\mu_0$.
The system of equations turns out to have no solution at all, which proves
that the slowly rotating Wahlquist solution cannot be matched to an
asymptotically flat vacuum exterior.

\section{Conclusions}
  The physical reason that the Wahlquist metric cannot describe an isolated rotating fluid body in equilibrium is apparently that this fluid medium would not withold the hydrostatic and gravitational forces acting in the neighborhood of the junction surface, without any deformation. To achieve equilibrium, additional, external forces are needed. There may exist a rotating fluid geometry with more degrees of freedom where all the conditions of matching can be met. In fact, in our approximation where we keep quadratic terms in the angular velocity, the perturbations  $h_0$ and $m_2-k_2$ are governed by uncoupled linear ordinary differential equations, of which the Wahlquist metric represents a particular solution. It is not difficult to obtain the {\it general} solution for the function $h_0$. We find, however, that this 'generalized' Wahlquist metric [with $m_2-k_2$ as given in Eqs. (\ref{secord})] still fails to satisfy the matching conditions. There is no {\sl a priory} reason that an initially spherically symmetric perfect fluid body could not be set in a rotational stationary equilibrium state. It would be most surprizing if the general rigidly rotating Whittaker metric could be shown impossible to match with the asymptotically flat vacuum exterior. Therefore, it will be desirable to carry out a check that this is not indeed the case. 

  Even though we have shown the impossibility of the Wahlquist metric describing an isolated rotating body in equilibrium, conceivably there may exist configurations where a Wahlquist fluid ball in an ambient vacuum domain is kept in equilibrium by an external force. A simple picture of this sort is when a fluid ball is surrounded by nested shells of vacuum and matter. It is our intention to investigate the stability of such more elaborate configurations in a perturbative treatment.

\section{Acknowledgments}
This work has been partially supported by the OTKA grant T022533. G.F. wishes to acknowledge the hospitality of the Department of Plasma Physics of Ume{\aa} University. Z.P. thanks the Wenner-Gren foundation for financial support. M.B. was partially supported by the NFR.


\begin{thebibliography}{9}
\bibitem{Lake}  M. S. R. Delgaty and K. Lake, Comput. Phys. Commun. {\bf 115} 395 (1998), gr-qc/9809013

\bibitem{Wahl}  H. D. Wahlquist, {\it Phys. Rev.} {\bf 172,} 1291 (1968). We use a tilde notation for Wahlquist's functions

\bibitem{Hoenselaers} C. Hoenselaers, in {\it Proc. Fourth Monash General Relativity 
Workshop}, Eds. A. Lun, L. Brewin and E. Chow, Monash Univ., (1993)

\bibitem{Whit}  J. M. Whittaker, {\it Proc. Roy. Soc. }{\bf A 306,} 1 (1968)

\bibitem{BrCo}  D. R. Brill and J. M. Cohen, {\it Phys. Rev. }{\bf 143.}
1011 (1966)

\bibitem{Hart}  J. B. Hartle, {\it Astrophys. J. }{\bf 150, }1005 (1967)

\bibitem{MS}  M. Mars and J. M. M. Senovilla, {\it Mod.Phys.Lett.A,} {\bf13},1509 (1998)

\bibitem{Roos}  W. Roos, {\it Gen. Rel. Grav. }{\bf 7, }431 (1976)
\end{thebibliography}
\end{document}